# Delegated Proof of Reputation:
# a novel Blockchain consensus

Thuat Do
Dept. Mathematics, Hong Kong
University of Science and Technology
Clear Water Bay, NT
Hong Kong
+84-392537614
thuat86@gmail.com

Thao Nguyen
Umbala INC
Ho Chi Minh City
Vietnam
+84-2836362936
thao@umbala.network

Hung Pham
Umbala INC
Ho Chi Minh City
Vietnam
+84-2836362936
ryanhungpham@umbala.network

## ABSTRACT
Consensus mechanism is the heart of any blockchain network. Many projects have proposed alternative protocols to improve restricted scalability of Proof of Work originated since Bitcoin. As an improvement of Delegated Proof of Stake, in this paper, we introduce a novel consensus, namely, Delegated Proof of Reputation, which is scalable, secure with an acceptable decentralization. Our innovative idea is replacing pure coin-staking by a reputation ranking system essentially based on ranking theories (PageRank, NCDawareRank and HodgeRank).

## 1. INTRODUCTION
Together with bitcoin, *Proof of Work* (PoW) has been introduced by Satoshi Nakamoto [6] since 2008. After that, Ethereum and some other blockchains use PoW protocol as well. People believe that such consensus helps the blockchains to be trustless and secure. Unfortunately, small networks are easily vulnerable by 51% attack [1]. Huge ones like Bitcoin and Ethereum still be possibly threatened as miners are centralized into several giant mining pools. In addition, PoW algorithm requires powerful computers to do intensive mathematical computation, hence is energy-inefficient. Another big issue of PoW is restricted scalability. Average numbers of transactions per second (TPS) on Bitcoin and Ethereum are 7 and 15, respectively. This is too slow for mass adoption. When those networks are busy, several individual transactions may take hours to days to be completed, and transaction fee possibly goes up much higher than usual.

Ones proposes *Proof of Stake* (PoS) as an alternative to PoW. The idea is that instead of possessing expensive and powerful hardware for mining tasks, ones are required to hold or stake (at least) a certain number of coins. Then the network randomly chooses someone to be a block producer. This solution is obviously energy-saving but the *long-range attack* and the *nothing at stake* problems arise on a PoS system. Furthermore, it may not scale greatly if all token holders are called for verifying transactions and approving blocks. To solve those issues, several projects, for instances, EOS, Bitshares, use PoS on their blockchain consensus with a modification so-called *Delegated Proof of Stake* (DPoS). Some authors criticize decentralization of such blockchains (as pseudo decentralized models). October 2018, EOS faced a corrupt governance – "mutual voting scandal" [2].

Vitalik Buterin, father of Ethereum, raises a *Blockchain Trilemma* that says about a trade-off among "security, decentralization and scalability". While security and decentralization are successfully achieved on several existing PoW blockchains (Bitcoin, Ethereum), scalability is still the most difficult problem.

Blockchain and distributed ledger technologies are in a very beginning stage but have a fantastic potential of application. Many projects are paying huge effort to build better public blockchains[2], even Ethereum is in transition from a pure PoW system to a hybrid (PoW+PoS) one to speed up TPS.

In this research, instead of solving the Trilemma, we focus on balance among three components: "placing security at first, improving scalability and offering an acceptable decentralization". In the next section, we are going to describe DPoS in more details, analyze its pros and cons. In Section 3, we introduce Delegated Proof of Reputation - our novel consensus mechanism as an improvement of DPoS. Therein, PageRank, NCDawareRank and HodgeRank theories is studied to make a modelling framework for our targets. Assessment and comparison with similar ideas of other projects are given as well. Section 4 discusses, among others, how innovative our solution is, and how it helps resolving vote buying problem happened on EOS and other issues of existing consensus protocols.

## 2. DELEGATED PROOF OF STAKE
In Peer-to-Peer networks like Bitcoin, all nodes have the equivalent role for transaction verification and block producing. It is a direct democracy. PoS is of analogous model. However, no pure PoS-based network has been launched up to now[3]. The reason may be the difficulty in solving security problems mentioned in the previous section.

Delegated Proof of Stake (DPoS) is operated as a representative democracy wherein stake holders vote for a small number of witnesses to secure and process transactions on the network. Here, without loss of general idea, we present DPoS on EOS. A token holder must stake coins in order to vote up delegates. More coins are staked, more votes are counted. Staked coins are locked in smartcontracts during voting rounds. Top-21 voted-delegates will be the block producers (BPs) who validate transactions, create new blocks and maintain the network, i.e. be responsible for the whole network operation. A block producer (BP) gets rewards for his work. Other delegates in Top-72 receive little rewards as well to serve as standby BPs. A BP may be voted off if missing his turn or be punished due to any compromise to the network. The punishment (possibly freezing or confiscating) is executed on the number of staked coins in his smartcontract. Thus, DPoS mechanism helps solving fundamental problems (the nothing at stake, the long-rang attack and weak subjectivity) of a naive PoS system. The rationales of DPoS essentially differ from PoW.

---

[1] Verge and Bitcoin-Gold suffered 51% attack with multi-million-dollar loss in 2018.

[2] i.e. Blockchains those are not operated by a single body or a group but open for everyone to join the networks.

[3] Silvio Micali, an MIT professor, and his team propose a pure PoS project (Algorand). They believe that it can solve the Blockchain Trilemma. https://www.algorand.com

- Token holders have a partially control on the network by their votes, and a chance to earn dividends. In contrary, PoW systems reward miners only. Miners may not own any coin, so they try best to maximize their profits. This makes a conflict between coin owners and mining bodies.
- DPoS significantly reduces the cost of network operation and maintenance while maximizes performance of blockchains, particularly scalability.

Several DPoS-based blockchains have successfully scaled up to thousands of TPS, for examples, EOS, WAVES, Steem. Naturally, PoS is energy-saving but still faces issues from centralized staking. Especially, DPoS model allows a small number of rich guys to control the network so it is not a fully decentralized model. Beside significant advantages, DPoS has several issues.

1. *Less incentive for standby BPs and voters*. In general, DPoS system rewards the vast amount for BPs, a little for standby ones and nothing for voters. Votes from token holders are important to maintain a partial decentralization of the model. With little reward, standby BPs still pay as equivalent infrastructure cost as BPs to wait for a lucky opportunity.

2. *More centralized approach*. The opportunity for a standby one to become a BP is small. In fact, the list of BPs (Top-21 delegates) of EOS network is almost unchanged for a long time (see "mutual voting scandal" [2]). Despite of voting, there is lack of diversity on the BP nodes as the whole network operation is in control of few richest bodies. This contradicts the decentralization philosophy of public blockchains.

3. *Common voters have lack of knowledge to assess BPs' performance*. Proof of Stake bases on the idea that token (money) owners have the right and responsibility. More money they own, more responsibility they should pay. However, this isn't always true in reality. There are many possible cases in which BPs commit bad actions. Since all delegates are public, BPs may collude to compromise the security of the network. Beside richness, what are supplement criteria that help voters choose right delegates?

4. *Application developers have no control on a blockchain, although their business is running on its top.* An application may bring thousands of transactions valued million dollars, but the developer doesn't have any control on the network without a significant staked amount. This conflict between top token holders and business owners is analogous to the one between miners and coin holders in PoW systems. BPs possibly compromise the network operating multi-million-valued applications. In that case, the developers have no safeguard unless they are in the group of BPs.

The fourth disadvantage can be generally considered as the conflict between value makers (workers) and money (value) holders. In a PoW system, coin holders have no control. In a DPoS system, workers don't have any right of operation (transaction verification, validation, etc), while they create the most important value for the network.

**Improvement approach**: our idea is to integrate working and staking into a consensus mechanism, giving proportional right of control and operation to application owners and token holders.

Fixing security as priority, we study how to balance decentralization and scalability in a model that is expected to satisfy all parties on a blockchain network. Then our targets are a high throughput blockchain and an acceptable semi-decentralization system. We are going to present our novel consensus mechanism, namely Delegated Proof of Reputation, in the next section that can solve the current issues of DPoS while inherits all its advances.

## 3. DELEGATED PROOF OF REPUTATION

As proposed in the last section, we introduce working (or performance) factors to modification of DPoS mechanism. Application developers are important contributors to the value of a blockchain, since they run their business on its top. They deserve to become network operators. We mean purely staking is not the only way to become a delegate. We replace stake voting by reputation voting. What is reputation?

### 3.1 Reputation score engine

Our reputation concept is a combination of three factors: staked amount, resource usage and transaction activity. A node or wallet may govern many addresses. In our voting mechanism, a node must register an official address (account) for rounds of reputation rating and voting. Thus, in this paper, for simplicity, we use the terms "node, address, account" alternatively without misunderstanding. Instead of pure stake voting, reputation score determines vote weight of an account. Let see our model and explanation in the following.

$$Vote\ weight = Reputation\ score = Rep = f(P, U, R)$$

$$Rep = w_1 P + w_2 U + w_3 R$$

**P** presents *stake power* of an account computed based-on the number of staked tokens.

**U** indicates (CPU, RAM, bandwidth) *resource usage* of an account. Such hardware component is indispensable on any internet-based network, hence it should be evaluated.

**R** is the most significant factor in our innovation. It stands for *transaction-based reputation rating* of an account. We appreciate active accounts with many (in/out) transactions. Like web-hyperlinks, transactions can be considered as a useful indicator for performance of an account and the entire network as well. It seems no benefit (additional values) for the network if token holders just stake to get blocking rewards in DPoS model.

$w_1, w_2, w_3$ are weights adding up to 1.

### 3.2 Stake power

Unlikely DPoS, we don't convert a staked amount instantly to the whole vote weight. If an account has an amount to be used for voting, it will be moved 10% to a staking smartcontract every day, then converted to stake power.

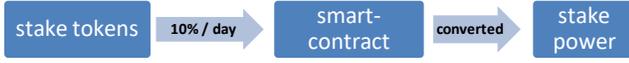

Assume that an account initially stakes $S$ tokens. To avoid infinite division, we require a common *threshold* $\theta$ as the minimum of tokens (for all stake amounts) to be sent to staking smartcontract.

$$\begin{cases} b_n = (1 - 0.9^{n-1})S, n \geq 1, \\ a_n = b_n \text{ if } S - b_n \geq \theta, \\ a_n = S \text{ if } S - b_n < \theta, \\ P_n = \dfrac{a_n}{A_n}, \end{cases}$$

where $a_n$ is the number of tokens stored in an account's staking smartcontract (i.e. convertible amount, accumulatively added 10% of $S$ each day), $A_n$ is the total sum of tokens in all staking smartcontracts in the entire network, and $P_n$ is stake power of the account, all parameters are computed at *n*-th day. Note that when the number of stake tokens is smaller than the threshold, it will send all of them to the smartcontract, making a full convertible amount as the initial stake. Fig 1 is an illustration for the case $S = 1000, \theta = 100, S - b_n < \theta \Leftrightarrow d > 21.8$. The convertible amount is full, i.e. $a_n = S = 1000$ at $d = 22$ afterward.

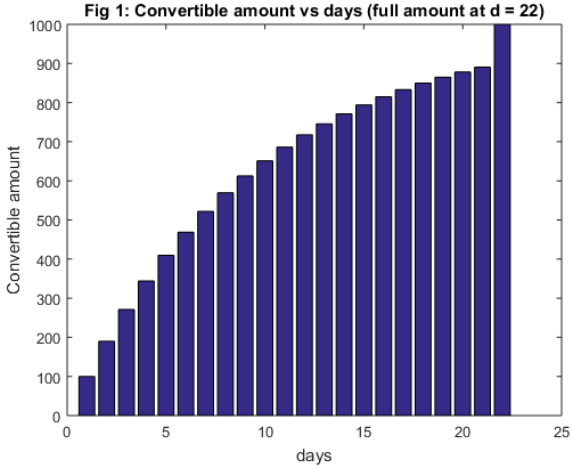

Why does it need days to be converted to full stake power? We think that time is necessary for accounts to show their behaviors. If somebody stakes a huge number of tokens for bad purposes, they cannot achieve in few days because our stake converter doesn't allow a strong power in a short time. For a fixed common threshold, larger amounts need more days to reach full. On the other hand, the converter encourages long-run accounts by accumulative stake power mechanism. Note that in our procedure, an account owner can increase his stake tokens anytime but the threshold and 10% converting proportion are fixed for the entire network as a common agreement.

Since token price changes according to market demand and sentiment, one can set a fiat-equivalence for the threshold, for example, $\theta \approx \$100 \pm 10$. Once the upper or under margins are broken, the minimum token threshold will change.

### 3.3 Resource usage

Computer and internet infrastructure are indispensable to maintain any blockchain network. Thus, we evaluate CPU, RAM, storage and bandwidth usage. We set the optimal usage (weighted as 1) from 68% to 88% the available processing resource. More concrete, we have the following framework (*x* indicates the average resource usage ratio within one day).

- $0 \leq x < 0.68$, weight increases as *x*. This promotes transactions when the system is in leisure status.
- $0.68 \leq x \leq 0.88$, weight equals 1. This presents the optimal resource usage.
- $0.88 < x \leq 1$, weight decreases as x. This discourages activities when the system is busy.

In this paper, we don't give explicit mathematical formulation for resource usage evaluation. However, it is not a difficult task and will be given in the future work together with experiments (see Section 5).

### 3.4 Reputation ranking

In Section 2, we discussed the lack of knowledge when common voters assess the delegates. Staked amount (by DPoS mechanism) is insufficient. How well do the delegates perform over time? This question motivates us to introduce a performance factor to DPoS consensus.

We believe that connection and linkages make value of any network. A node within a network is important somehow if there are linkages that refer to it. More referring linkages, more important a node seems to be. This is the essential idea of PageRank [3]. Among thousands of websites, relative importance of a site is evaluated by hyperlinks referring to it. PageRank mathematically models web pages as vertices and hyperlinks as edges of a directed graph. Various algorithms to rank web pages are proposed, and they are a core part of a search engine (for instance, Google Search). As an efficient alternative of PageRank, NCDawareRank [1] was introduced in 2013 which is resistant to link spamming and gives a block-structure approach to the web graph. Another ranking theory that can be applied here is HodgeRank [7], which exploits the graph Helmholtzian to decompose every edge flow into two orthogonal components: a gradient flow (representing a global ranking) and a divergence-free flow (validating the global rank). NEM project [4] introduced Proof of Importance (POI) based on NCDawareRank to rate node performance. However, we find limitations in NEM ranking model as well as application of NCDawareRank to rating problem on nodes of a blockchain network. The details are discussed in next subsections. Comparison and suitability assessment among three mentioned theories will help us to choose the best choice for our reputation ranking system. Then, we will build a statistical ranking model for a blockchain network.

*3.4.1 Web ranking theories: a comparison*

Assume the we have a set of *n* web pages with an out-link matrix $L = (L_{uv})$ where $L_{uv}$ is the number of links from page *u* to page *v*. Let $d_u = \sum_v L_{uv}$ is the out-degree of u. We normalize to have a stochastic matrix $A = (A_{uv})$ where $A_{uv} = \dfrac{1}{d_u}$ if there is a link from *u* to *v*, and $A_{uv} = 0$ otherwise. Page rank is defined to be a vector *R* that assigns rating values over the pages based on the out-link matrix. In a simple way, $R = cAR$, i.e. *R* is an eigenvector (with unit-$L_1$ norm) corresponding to the largest eigenvalue *c*. However, to prevent the effect of dangling pages (i.e. the pages with no out-link), the *damping factor* $\alpha$ and a *teleportation* matrix *E* are introduced. Let *p* be the probability vector that initially assign an even rating (1/*n*) to all pages, and $E = ep^T$ where *e* is the identity matrix. Then we obtain a PageRank model [3]

$$H = \alpha A + (1 - \alpha)E, R = cHR.$$

As a generalization of PageRank, NCDawareRank [1] introduce a new term, the *inter-level proximity* matrix *P*, that exploits an important property that the Web can be partitioned into nearly completely decomposable (NCD) blocks. Let $\{B_i\}_1^k$ be an NCD-block partition on the web graph. Any page $u$ is contained in a unique block $B_i$, we denote $B(u)$ for convenience. Let $G_u$ be the set of pages that $u$ refers to, and $X_u$ be the set of proximal pages of $u$, i.e. $X_u = \bigcup_{v \in \{u \cup G_u\}} B(v)$. Then $P_{uv} = \frac{1}{d_u|B(v)|}$ if $v \in X_u$ and $P_{uv} = 0$ otherwise. The resulting matrix is

$$Q = \alpha A + \mu P + (1 - \alpha - \mu)E, R = cQR.$$

In addition, $M$ can be factorized into two extremely sparse matrices, making NCDawareRank computed more efficiently. Further, block-structure approach of NCDawareRank is the intuition to prevent link spamming. However, scanning proximal nodes is a heavy task, takes long time to complete.

HogeRank [7] approaches the rating problem in a distinguished and general way. Using combinatorial Hodge theory, it exploits the graph Helmholtzian to provide a beautiful ranking theory. HodgeRank views edge flows as pairwise rankings, then decomposes it into three orthogonal components as followed.

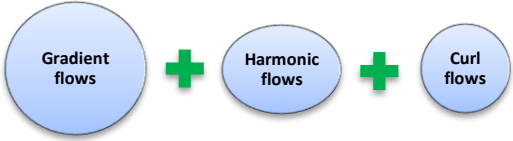

The first component gives us a global ranking and the sum of the last two terms tells us how good the rating is.

Before going to interpretation in details, we need some preliminaries. A Markov chain on the pages is given by $M_{uv} = \frac{\alpha L_{uv}}{d_u} + \frac{1-\alpha}{n}$ (PageRank model). Then an edge flow is $Y_{uv} = \log \frac{M_{uv}}{M_{vu}}$, and weight $w_{uv} = 1$ if $L_{uv} + L_{vu} > 0$ and $w_{uv} = 0$ otherwise. Therefore, $Y = (Y_{uv})$ is skew symmetric, and $W = (w_{uv})$ is a symmetric $\{0, 1\}$-valued matrix. HodgeRank [7] aims to find a rank-2 skew symmetric matrix $X$ that approximates $Y$ and satisfies $X_{uv} = s_v - s_u$, where $V$ is the set of vertices (nodes) and $s: V \to \mathbb{R}$ is a real-valued function. Thus, $X$ induces a global ranking via the rule $u \leqslant v$ if and only if $s_u \leq s_v$. Furthermore, $s$ is the unique score function (up to an additive constant) that assigns each node $u$ a rating score $s_u = s(u)$, hence the ranking vector $R = s(V)$. Matrix $X$ is the solution of the least square problem

$$\operatorname{argmin}_X \sum_{ij} w_{ij}(X_{ij} - Y_{ij})^2 \quad (3.4.1)$$

HodgeRank doesn't need the proximity matrix. In the view of computation complexity, the least square problem of HodgeRank is less expensive than eigenvector computing proposed by PageRank and NCDawareRank. We are going to get a deeper insight of the Hodge decomposition.

Let $F = \left\{\{i,j\} \in \binom{V}{2} : w_{ij} > 0\right\}$ and $T(F) = \Big\{\{i,j,k\} \in \binom{V}{3} : \{i,j\},\{j,k\},\{k,i\} \in F\Big\}$. HodgeRank theory formally defines a pairwise ranking edge flow $X: V \times V \to \mathbb{R}$ to be:

1. *consistent* on $\{i, j, k\} \in T(F)$ if it is curl-free on $\{i, j, k\}$, i.e. $(\operatorname{curl} X)(i, j, k) = X_{ij} + X_{jk} + X_{ki} = 0$;
2. *globally consistent* if it is a gradient of a score function, i.e. $X = \operatorname{grad} s = s_j - s_i$ for some $s: V \to \mathbb{R}$;
3. *locally consistent* or *triangularly consistent* if it is curl-free on every triangle in $T(F)$;
4. *cyclic ranking* if it contains any inconsistency, i.e. there exist nodes $i, j, k, \ldots, p, q \in V$ such that

$$X_{ij} + X_{jk} + \cdots + X_{pq} + X_{qi} \neq 0.$$

**Hodge decomposition.**

*Edge flows = (Gradient $\oplus$ Harmonic $\oplus$ Curl) flows.*

*Divergence-free flows = Harmonic flows $\oplus$ Curl flows.*

The gradient flows comprise the *globally consistent* (or *acyclic*) pairwise rankings which induce a unique rating score function on the nodes. The divergence-free flows (in which in-flows and out-flows of each node are equal) are interpreted as *cyclic* rankings (i.e. of the form $i \leqslant j \leqslant k \leqslant \cdots \leqslant i$) and of course they are inconsistent. The sum of the gradient and harmonic flows is the subspace of curl-free on every triangle and it is locally consistent. The harmonic component is both curl-free and divergence-free, is locally consistent but not globally. The last term is locally cyclic pairwise rankings. Naturally, our greatest concern is the gradient flows which induce a globally consistent ranking. However, divergence-free flows are useful for validation. If it is small (measured by a norm) compared to the gradient flows, then our ranking vector and rating scores are good and reliable. Furthermore, harmonic flows and curl flows can give us some insight to apply penalty on spamming nodes.

HodgeRank is not only a good approximation to the PageRank stationary distribution but also a useful technique to study the inconsistency in PageRank Markov model. In the next subsection, we will discuss in more details to see that HodgeRank theory is the best choice for our ranking problem on blockchain nodes. Now, we sum up a theoretic comparison among the three mentioned ranking theories.

|  | PageRank | NCDaware-Rank | HodgeRank |
|---|---|---|---|
| Proximity matrix | NO | YES | NO |
| Complexity rate | X | X > Y | Y > Z |
| Validation | NO | NO | YES |
| Spamming resistance | NO | YES | YES |

### 3.4.2 A graph model of a blockchain network

There are three essential distinctions between the Web and a blockchain network.

1. Hyperlinks carry no value while transactions transfer cryptocurrency.
2. No matter one or many links from a page A to a page B, adjacency matrix of PageRank and NCDawareRank models represents that linkage as 1. Between two accounts in a blockchain network, there are possibly many transactions. They are all worthy so should not be excluded.
3. Within a website, subpages usually have a dense internal interaction while much fewer external links (i.e. with other websites). This leads researchers to study block structure of the Web graph. In contrary, a (normal) user may govern multiple accounts (or addresses) on a blockchain but internal transferring is rare, while transactions are usually executed with others. Thus, block structure approach is appropriate for the Web but not precisely suitable for blockchains.

Our target is to make a mathematical model that can represent a blockchain with all essential characteristics. We treat an address as a vertex in the graph. Each transaction is an edge connecting two addresses (2 vertices), distinguished by in/out transfer, i.e. making the graph directed. Assume that we are computing reputation ranking score with the following assumptions.

- There are $n$ consecutive blocks $\{B_k\}_1^n$ collected within $d$ days, $B_1$ is the oldest block and $B_n$ is the latest one.
- The sum of all transferred tokens in each block $B_k$ is $C_k$ and the total is $C = C_1 + C_2 + \cdots + C_n$.
- The number of all transactions in each block $B_k$ is $T_k$ and the total is $T = T_1 + T_2 + \cdots + T_n$.
- There are $m$ accounts $\{A_i\}_1^m$ with at least one transaction (in fact, a block doesn't contain any address without any transaction).

We remind readers that blockchain network can be considered as internet of value. Therefore, we take transaction amounts into account as priority. Contrasting to NEM [4], we don't set a minimum of tokens in each transaction to be counted for rating rounds. This, together with zero-fee transaction, makes our blockchain model more suitable for micro-payment (an essential expected-application of Blockchain and cryptocurrency). One more distinction to NEM, we still calculate reputation rating of accounts with zero stake power (i.e. without staking). We consider them as popular users on the network. They deserve to have some reputation and they can use it to vote for delegates.

Regarding each block, we apply a single rule, i.e. if there are many transfers from an account to another, we count as a single transaction with added-up total of tokens. Assume that at block $B_k$, account $A_i$ transfers $x_k$ tokens to account $A_j$, then we calculate the weighted transaction as followed

$$w_{ijk} = x_k \varphi^{\left(-\eta\left\lceil k\frac{d}{n}\right\rceil\right)}, \quad (3.4.2)$$

where $\lceil z \rceil$ is the ceiling function mapping the least integer greater than or equal to $z$, hence $\left\lceil k\frac{d}{n} \right\rceil$ is the birthday of $k$-th block. The second term assigns preference to later block with a base $\varphi > 1$ and a shrinkage coefficient $\eta \geq 0$ fixed for every block and rating round (in NEM, $\varphi = e \approx 2.72, -\eta = \ln 0.9$). Shrinkage is useful because recent activities have more meaning to the current status of the network than very old transactions (for instance, happened months or years ago). A transaction amount weight decreases to nearly zero if the number of days is large enough. Fig 2 shows how an amount of 1000 weighted over days, $\varphi = 2, \eta = 0.02, n = 10000$. If $d$ is small, for instance, $d < 10$ (days), ones can ignore the exponential term.

Varying $n$ blocks, the total weighted (out) transfers from $A_i$ to $A_j$ is

$$W_{ij} = \sum_{k=1}^n w_{ijk} = \sum_{k=1}^n x_k \varphi^{\left(-\eta\left\lceil k\frac{d}{n}\right\rceil\right)}.$$

The equation guarantees that transactions with zero token worth nothing in the total weights. In NEM model [4], the outlink matrix $O$ is given by positive net flow only, i.e.

$$O_{ij} = \begin{cases} \frac{\widetilde{w}_{ij}}{\sum_i \widetilde{w}_{ij}} & \text{if } \sum_i \widetilde{w}_{ij} > 0 \\ 0 & \text{otherwise} \end{cases} \text{ where } \widetilde{w}_{ij} = \begin{cases} W_{ji} - W_{ij} > 0 \\ 0 \text{ otherwise} \end{cases}.$$

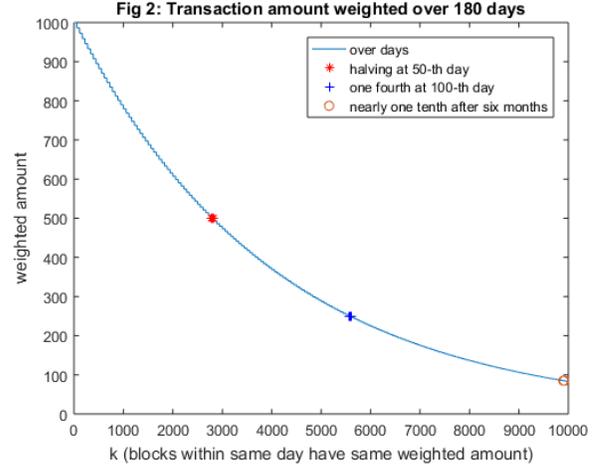

Fig 2: Transaction amount weighted over 180 days

We have a different mindset that every transaction is meaningful somehow, it should be preserved in the mathematical model. Therefore, our (weighted) out-transfer matrix is

$$L = (L_{ij}) = \left(\frac{W_{ij}}{C}\right),$$

normalized by the total transferred tokens within a rating round. If the duration of block collecting is long, for examples, several months, ones can apply shrinkage on $C_k$ to obtain another denominator for normalization (which assigns higher preference to later transaction volumes) in the out-transfer matrix.

$$\hat{C} = \sum_k C_k \varphi^{\left(-\eta\left\lceil k\frac{d}{n}\right\rceil\right)}, \text{ hence } L = (L_{ij}) = \left(\frac{W_{ij}}{\hat{C}}\right).$$

If the shrinkage is used, it should be applied in computing both the out-transfer weights and the out-transfer matrix.

Now, we are ready to construct a Markov chain and an edge flow on the nodes by

$$M_{ij} = \frac{\alpha L_{ij}}{d_i} + \frac{1-\alpha}{n} \quad \text{and} \quad Y_{ij} = \log\frac{M_{ij}}{M_{ji}}. \quad (3.4.3)$$

Note that our edge flow model exploits the reversible property of a Markov chain (captured in PageRank model).

$$\pi_i X_{ij} = \pi_j X_{ji}$$

According to [7], a reversible Markov chain $X$ has a pairwise ranking flow

$$X_{ij} = \log\frac{M_{ij}}{M_{ji}} = \log \pi_j - \log \pi_i,$$

where $\pi$ is a stationary distribution, and $\log \pi$ gives the global ranking. Therefore, we aim to find

$$X^* = \operatorname{argmin}_X \|X_{ij} - Y_{ij}\|_2,$$

which is compatible with Equation (3.4.1). This implies that the Hodge decomposition of edge flows gives us the best reversible approximation of the Markov chain, and $X^*$ is exactly the gradient flows that induces a globally consistent ranking and a rating score function.

Note that in our mathematical model, the quantity of transactions has evaluated yet. A compose of transferring amount and transaction quantity is expected in the next research.

## 3.5 Reputation score

Assume that from the gradient flow in Hodge decomposition, we obtain a ranking vector that assigns rating scores $R$ over the nodes. Together with stake power ($P$) and resource usage ($U$), we compute the final reputation scores (Rep) of an account (a node) as followed

$$\text{Rep} = w_1 P + w_2 U + w_3 R, \quad (3.4.4)$$
$$\text{where } w_1 = 0.4, w_2 = w_3 = 0.3.$$

The focal point of our consensus mechanism is a comprehensive balance between benefit and responsibility over all network participants. We believe that our reputation ranking system is fair among stake holders, resource contributors and application developers. Almost accounts on the network have positive reputation scores without any limitation or eligibility condition. Stake power enables token holders to gain vote weight but not instantly strong. They need long-run to reach maximal power. Resource usage reflects the hardware and internet infrastructure that is crucial to operate and maintain a blockchain. Reputation scores present the transaction and business activities, the vital characteristic of a network.

## 4. FURTHER DISCUSSION

Firstly, our Delegated Proof of Reputation (DpoR) can be considered as "*semi-decentralization*", *decentralization* in the sense of reputation voting and *semi* in the side of delegated block producers. Distributing reputation ranking to almost network participants makes a great diversity on voters, delegates and BPs (greater than DPoS of EOS or POI of Nem). Therefore, no specific body or group has a dominate control on the network.

Unlikely instant stake counting in DPoS, full stake power applies for long-runners in DPoR. Staked amount needs days to be converted to stake power, meanwhile the holders must show their honesty to be voted. Thus, rich guys cannot gain an immediately strong impact on the network.

It is the first time that application developers are granted a controlling opportunity on a blockchain. They make the main value of a network via their business and transactions. That why we invent proof of reputation as a sophisticated balance between the two most important groups: application developers (value makers) and token holders (value owners). We also believe that running business is more important than holding money. Thus, we assign a 60% weight to resource usage and reputation ranking, greater than 40% of stake power. This incentivizes developers to build their business on the network and helps preventing malicious rich stake holders.

Furthermore, our converter helps reducing the staking vs cash flow conflict. Cash flow measures health of an economy and should be promoted. In contrast, DPoS requires a significant amount for staking, so reduces token circulation. In our DPoR, staking is not the dominate factor, hence it doesn't affect negatively to cash flow. Transaction-based rating, on the other side, is an incentive for token flows to move more actively, then improves liquidity of the economy.

Note that our reputation score engine is not real-time running. It is computed periodically after a long period of time (for example, hours or days), hence doesn't affect the processing efficiency of the nodes and the entire network. Moreover, nodes without powerful processors may refer to the reputation scores computed by others or separately independent machines of a third audit party, for instance, a statistical analytic firm.

**Table: a simple comparison** (Med means Medium)

|  | PoW | PoS | DPoS | POI | DPoR |
|---|---|---|---|---|---|
| Energy saving | No | Yes | Yes | Yes | Yes |
| Decentralization | Full | Full | Pseudo | Med | Semi |
| Scalability | Low | Med | High | High | High |
| Staking | No | Yes | Yes | Yes | Yes |
| Reputation credit | No | No | No | Yes | Yes |
| Voting | No | No | Yes | No | Yes |
| Fairness & balance | Med | Med | Low | Med | High |
| Incentive for node honesty and responsibility | High | Low | Med | Med | High |

**Boosting token flow (loop attack) problem.** When staking doesn't bring a great control, dishonest nodes may transfer tokens around accounts under their governance many times (see Fig 3) in order to gain higher reputation score, since DPoR appreciates quantity and token amount of transactions. To fight against boosting, NEM proposed positive net flow for POI (i.e. net flow equals out-flow minus in-flow if the difference is positive, and zero otherwise). A loop, therefore, doesn't worth higher importance scores than transferring just once, despite of millions of transactions. However, NEM's convention has a limitation when ignores the case of negative difference. We believe that every transaction has some meaning, so we try to preserve it by another solution. Equation (3.4.3) constructs the transferring (edge flow) matrix by ratio between in-flow and out-flow, hence preserves all information and guarantees that multi-transactions in a loop isn't worthier than once (for example, 1000:1000 = 1:1). More important, the ratio keeps useful information for Hodge decomposition, then harmonic and curl flows show us loops and involved nodes. Accounts will refer to that token flows, assess node honesty and decide who they should vote for. Therefore, DPoR is resistant to boosting (and Sybil attack as well).

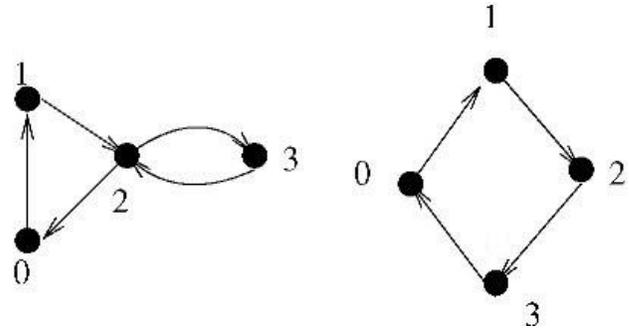

**Fig 3. Transfers in loops**

**Increasing computation difficulty over time**. The shrinkage in Equation (3.4.2) says that all transactions older than one year are weighted nearly zero, hence we can skip them. However, collected blocks within several months possibly contains thousands of accounts and multi-millions of transfers, while DPoR still computes old-months-transactions. This is waste of time and computation resource.

In this paper, we review several major blockchains, their consensus mechanisms together with advantages and disadvantages. Then we suggest innovation ideas for conventional DPoS. Ranking theories (PageRank, NCDawareRank, HodgeRank) are studied to introduce a novel consensus mechanism – Delegated Proof of Reputation. Analysis on ranking theories are given. A mathematical model is constructed for DPoR together with extensive discussion and comparison with others. However, within this research, we cannot give a comprehensive review on advances and limitations of DPoR. Theoretical framework is given but experiments are necessary to validate suitability of HodgeRank to DPoR and to assess the reputation score engine.

## 5. FUTURE WORK

We know that a lot of future investigations needed to be conducted. We are building a blockchain based on Delegated Proof of Reputation (see https://umbala.network/). Experiments are necessary to verify our innovative ideas and theoretical framework before running testnet.

Although DPoR is resistant to loop attack, we want a useful tool to prevent boosting intension. Millions of boosted transactions worth nothing while they increase block producers' workload and network's data volume. Transaction fee can discourage boosting flow. However, our project, Umbala Network, applies fee-less transaction strategy to serve micro-payment. A suggestion is creating a filter based on Hodge decomposition. Punishment (freezing or confiscating staking tokens) can be applied on nodes whose token flows are in harmonic and curl subspaces (i.e. with global inconsistency).

A solution to avoid repeating computation on the old blocks of the previous rating rounds is helpful. We mean an updatable algorithm for reputation score computing. Online HodgeRank [5] gives online preferential attachment sampling and online algorithms to update rating and track curl flows. How can we apply to blockchain model and the updating problem?

How can we add the transaction quantity into mathematical model of reputation ranking? A compose of transaction quantity and transferring amount may improve our model. More clearly, in Equation (3.4.2) and Equation (3.4.3), if we can use the number $T_k$ of all transactions in each block and the total $T = T_1 + T_2 + \cdots + T_n$ combining all considering blocks, then the model may present characteristics and activities on the blockchain better.

A partial reward for voters is a good incentive to maintain a high decentralization level in voting rounds. A more comprehensive assessment on DPoR and comparison with others is useful for blockchain application developers.

## 6. ACKNOWLEDGMENTS

# Columns on Last Page Should Be Made As Close As Possible to Equal Length

## Authors' background

| Your Name | Title* | Research Field | Personal website |
|---|---|---|---|
| Thuat Do | PhD candidate | Data Science, Blockchain | |
| Thao Nguyen | CEO, Umbala Network | | |
| Hung Pham | CTO, Umbala Network | | |

*This form helps us to understand your paper better, **the form itself will not be published.**

*Title can be chosen from: master student, Phd candidate, assistant professor, lecture, senior lecture, associate professor, full professor